\documentstyle[11pt,AATS,epsf]{article}	

\begin{document}	

\title{Type Ia Supernovae, the Hubble Constant, the Cosmological 
Constant, and the Age of the Universe} 

\author{John L. Tonry}
\affil{Institute for Astronomy, University of Hawaii}
\author{The High-Z Supernova Search Team}
\affil{http://cfa-www.harvard.edu/cfa/oir/Research/supernova/HighZ.html}


\begin{abstract}
The age of the Universe depends on both the present-day Hubble
Constant and on the history of cosmic expansion.  For decelerating
cosmologies such as $\Omega_m$= 1, the dimensionless product
$H_0\,t_0<1$ and modestly high values of the Hubble constant $H_0 > 70$
would be inconsistent with a cosmic age $t_0$ larger than 12 Gyr.  But
if $\Omega_\Lambda > 0$, then $H_0\,t_0$ can take on a range of values.
Evidence from the Hubble diagram for high redshift Type Ia supernovae
favors $\Omega_\Lambda\sim0.7$ and $H_0\,t_0 \sim 1$.  Then, if $H_0$
lies in the range 65--73, the age of the Universe, $t_0$, is $14\pm1.6$~Gyr.
\end{abstract}


\section{It has been an interesting five years!}

Five years ago, the
combination of deep seated belief in inflation, implying $\Omega = 1$,
and stellar age estimates near 15 Gyr seemed to require $H_0\sim40$.
Measurements of $H_0\sim 60$ and $\Omega_m\sim0.3$ in clusters
notwithstanding, Bartlett, Blanchard, Silk, and Turner wrote a
provocative paper entitled, ``The Case for a Hubble Constant of 30
km/s/Mpc.''  Persuaded by the power of
theoretical reasoning, Joe Silk bet Brian
Schmidt and me a case of Scotch that $H_0 < 60$.  While Joe has not
yet paid up, in the past 5 years he has moved closer to the source of
Scotch while the Hubble constant has moved to 60 and beyond.  The new
element is that supernovae have made the connection between $\Omega =
1$ and the cosmic age more flexible because of plausible evidence for
cosmic acceleration.
 
A Danish-English team (Norgaard-Neilsen et al. 1989) initiated a
program to find supernovae in clusters of galaxies at redshifts of
0.3-0.5, with the idea that they could distinguish the effects of
cosmic deceleration, as expected in an $\Omega_m$ = 1 universe by
measuring the peak apparent magnitudes of supernova light curves.  The
observational problem was to find these faint (m$\sim$21-22) and
distant supernovae near the peak of their light curves.  But small
detectors kept this pioneering effort from yielding significant
results.

The Supernova Cosmology Project (SCP) based at Lawrence
Berkeley Lab forged ahead with further attempts to find distant
supernovae by extending the methods of the Danes to bigger, faster
telescopes.  After abandoning attempts to instrument the AAT prime
focus for this purpose, they used the standard large format detectors
at the Kitt Peak National Observatory starting in 1992 (Perlmutter
1995).  By 1997, they had a preliminary result (Perlmutter et
al. 1997) based on observations of seven supernovae discovered in 1994
and 1995.  By comparing their supernovae with the sample at low
redshift, they concluded that the evidence favored a universe with
high matter density $\Omega_m = 0.88 \pm 0.6 $.  They argued that the
supernova data at that point placed the strongest constraint on the
possible value of the cosmological constant, with their best estimate
being $\Omega_\Lambda = 0.06$.

Since 1995, there have been two groups pursuing evidence on cosmic
deceleration using the Hubble diagram for supernovae.  Our High-Z
Supernova Search Team, steered by Brian Schmidt, and encompassing
workers on four continents and one mid-Pacific island, found our first
supernova in 1995 (Schmidt et al. 1998). From three additional supernovae
studied with $HST$ the High-Z Team found a low value
of $\Omega_m$, showing that ordinary matter could not close the Universe
(Garnavich et al. 1998).  Our first results on
$\Omega_\Lambda$ were reported at the Dark Matter conference in
February of 1998 by Filippenko and Riess (1998) and published in the
Astronomical Journal in September 1998 (Riess et al. 1998).  In the
same period, the SCP revised their analysis of earlier data
(Perlmutter et al. 1998) and then independently reported evidence on
$\Omega_\Lambda$ in the Astrophysical Journal in June 1999 (Perlmutter
et al. 1999).

\section{SNIa constraints on $t_0$}

Even this first round of supernova observations, which emphasized a
sample near z$\sim$0.5, provided a good constraint on the
{\it{difference}} $\Omega_m$-$\Omega_\Lambda$ which, since $\Omega_m$
measures deceleration, and $\Omega_\Lambda$ measures acceleration,
translates into a surprisingly tight constraint on $H_0\,t_0$.

The present samples of SNIa published, in hand, and being reduced by
the two teams provide a statistically robust measurement that
$\Omega_\Lambda > 0$.  The important questions now are whether
the supernovae at large redshift are really the same as the supernovae
nearby and whether exotic forms of grey dust might obscure both the
supernovae and our understanding of cosmology (Aguirre, 1999, 2000).
The observational approach to answer these questions is to use spectra
to examine the question of homogeneity (Coil et al. 2001) and
multicolor observations over a wide range of wavelengths (as might be
done with a superb 8-m infrared telescope at the world's best site) to
constrain the properties of intergalactic dust (Riess et al. 2000).
So far, although the distant supernovae could have failed these tests,
they seem indistinguishable from the SNIa nearby.
 
A more ambitious test for the cosmological origin of the observed
effect is to extend the data set to higher redshift.  While dimming
due to evolution or dust would most naturally lead to larger effects
at higher redshift, cosmological effects could have the opposite sign
due to cosmic deceleration at early epochs (z $\sim$ 1.5), followed by
a transition to acceleration in the more recent past.

\subsection{What does a supernova look like?}

Looking at real data helps develop an understanding of the
observational issues in discovering and measuring light curves for
high redshift supernovae.  Here we illustrate the appearance of a
z=0.81 supernova, SN 1999fj, as observed in a series of images from
October through December 1999. Large format detectors on telescopes at
good seeing sites are the chief requirement for efficient surveys to
find type Ia supernovae at $m\sim24.5$.  Follow-up observations with
8-m class telescopes enable us to construct light curves that can be
used to place each SNIa firmly on the Hubble diagram.
 
\begin{figure}[htpb!]	
\plotfiddle{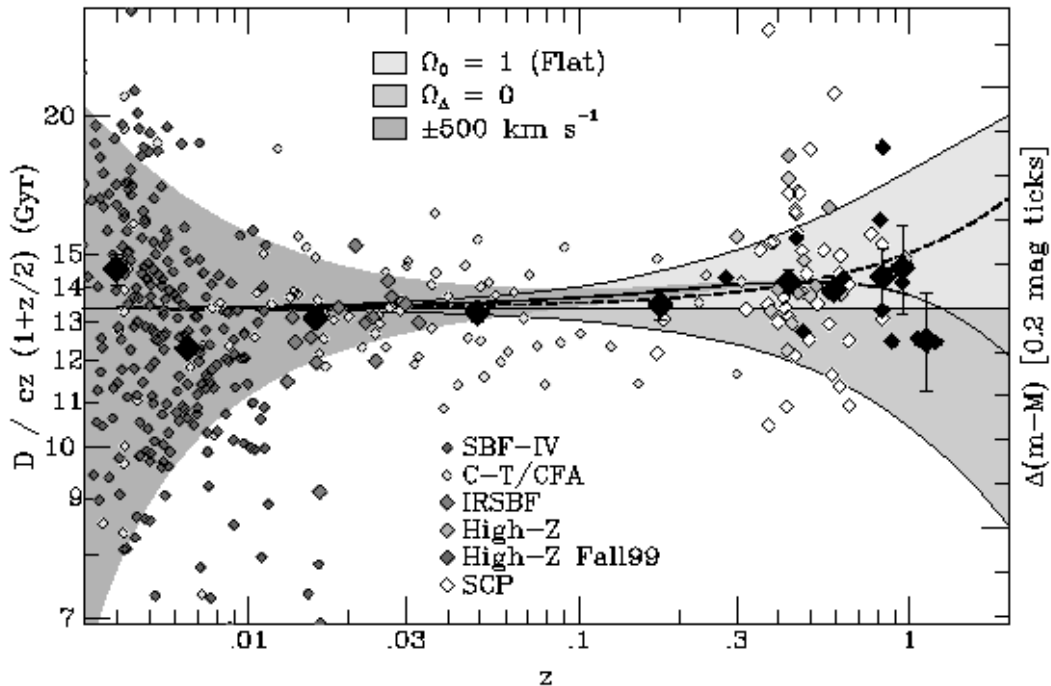}{320pt}{0}{70}{70}{-210pt}{-115pt}
\caption{SN 1999fj at four epochs. 03 Oct 99 has the supernova barely
above our detection threshold using $I$ band at the CFHT.  03 Nov 99 is
the discovery epoch.  The reader is invited to find the new dot.  This
image covers approximately 1/3000 the area of the detector array, so
the real search is automated.  15 Nov 99 shows an image in
0.5\arcsec\ seeing from the VLT, and 15 Dec 99 shows an image in 0.9\arcsec\
seeing from the Keck telescope. \label{natasha} }
\end{figure}

\subsection{Light curves from Fall 1999}

One of the key developments that makes SNIa so useful as standard
candles is the discovery by Mark Phillips (Phillips 1993) that the
luminosity of a SNIa is related to its rate of decline after maximum
light.  This approach, refined by Hamuy et al. (1996) and by Riess,
Press \& Kirshner (1996) allows the distance of a well-observed SNIa
with two color data to be determined to better than 10\%.  

The light curves for high redshift supernovae are obtained in filters
which can be transformed back to rest frame B and V with good
precision.  Time dilation, a signature of cosmic expansion, is a
powerful effect for supernovae near z$\sim 1$, transforming 40 days in
the observer's frame to 20 days in the supernova's own rest frame
(Leibundgut et al. 1995).  Photometric calibration and scrupulous
subtraction of galaxy light are serious problems for this work which
become more difficult at high redshift.  Nevertheless, images like
those in Figure 1 can be used to construct light curves as shown in
Figure 2.

\begin{figure}[htbp!]	
\plotfiddle{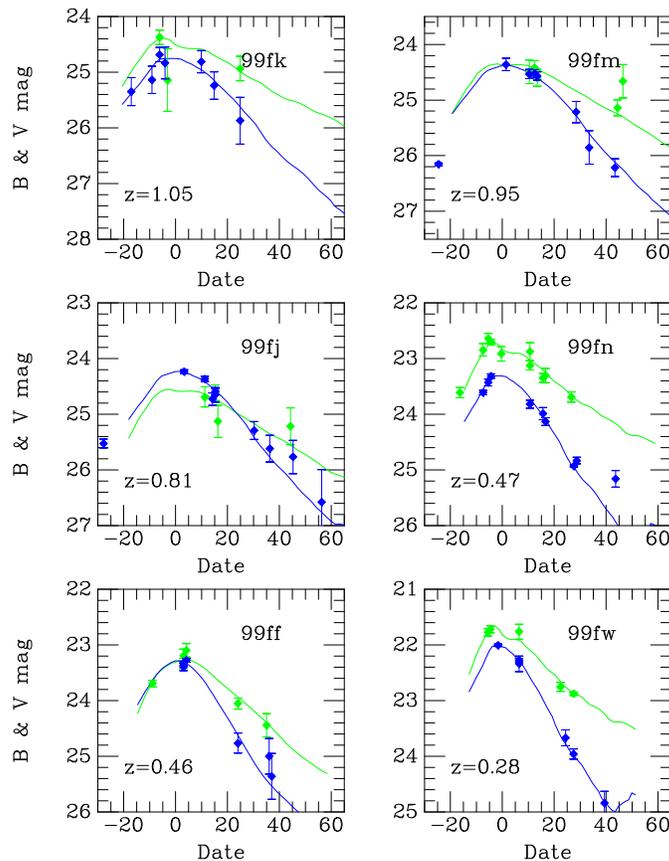}{310pt}{0}{50}{50}{-120pt}{-10pt}
\caption{A sampling of two-color light curves for 6 supernovae from
Fall 1999, ranging from z=1.05 to z=0.28 \label{lc} }
\end{figure}

\section{Meditations on $H_0$}

What is the observational evidence on $H_0$?  Many techniques have
been developed for measuring extragalactic distances and the ones we
love the best are SNIa (Jha et al. 1999) and the Surface
Brightness Fluctuation method, reviewed recently by Blakeslee, Ajhar,
\& Tonry (1999).  Ajhar et al. 2001 demonstrate that these methods are
internally consistent: as Figure 3 shows, distances measured by SNIa
and by SBF to the same galaxies are consistent within the quoted
errors.  This suggests that both methods are in good shape.  

But what
is $H_0$, so diligently sought through the decades?  This depends
entirely on the sample of Cepheids used and the distances assigned to
those Cepheids by expert workers in that field.  If you use the
distances to galaxies with Cepheids as determined by various papers by
the Key Project (Ferrarese et al. 2000, Freedman et al. 2001, Gibson
\& Stetson 2001) to define the absolute magnitude of SNIa and SBF, you
get $H_0 = 73$, 75, and 77 for both SNIa and SBF.  If you use the
distances to the same galaxies using the same Cepheid observations as
reduced by Saha et al (1997) and most recently compiled by Parodi et
al. (2000), you get $H_0 = 65$ according to the methodology used by
the High-Z team, or $H_0 = 58.5$ according to Parodi et al.
Similarly, the distances to the SBF calibrating galaxies depend
entirely on the Cepheid distances.  

The Hubble constant is {\bf not}
determined by SNIa or by SBF alone: these methods give excellent relative
distances to galaxies and tie the cosmic expansion firmly to the local
calibrators, but the calibration by Cepheids is now the largest
uncertainty in measuring the local value of the Hubble Constant.  For
the rest of this paper, we will either adopt $H_0=73$ as a best-guess
value, or else a probability distribution which consists of two
Gaussians of fractional width 0.1, centered on $H_0=73$ and $H_0=65$,
with a $2/3$, $1/3$ weighting.  It ain't perfect, but it's our gut
feeling of where $H_0$ really lies.

\begin{figure}[htbp!]	
\plotfiddle{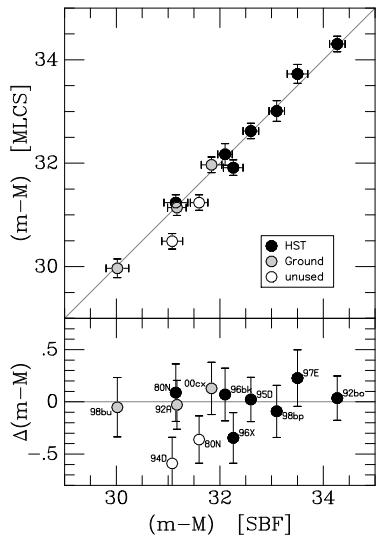}{180pt}{0}{30}{30}{-80pt}{-25pt}
\caption{There is no discrepancy between the SNIa and SBF when they
get their zero points from the same set of Cepheids.
\label{sbfsn} }
\end{figure}

\section{Differential Hubble diagram for SBF and SNIa}
 
\begin{figure}[htbp!]
\caption{The cosmological diagram
for SBF and supernovae, showing observed luminosity distance (units of Glyr)
divided by $cz$ and $1+z/2$ (i.e. divided by luminosity distance in
an empty universe) as a function of redshift.  Both sets of distances
are normalized to an $H_0 =73$ zeropoint;
the left hand intercept is therefore
$H_0^{-1} \sim t_0$; the right axis is labelled in magnitudes.
The darker gray region shows where
cosmological models with $\Omega_\Lambda = 0$ lie, and the light and dark
gray regions show where flat models ($\Omega_0 = 1$) lie.  The
black downturning curve shows an $(0.3,0.7)$ cosmology, enhanced by
the darkest gray region of $\pm500$~km/s which corresponds to $\pm2.5\sigma$
thermal peculiar velocities.  The large black points are
medians in various redshift bins, and the error bars are estimates of
the uncertainty in the median judged from the scatter of the
contributing points.  At $z\sim0.5$ these points lie
very significantly above the region permitted if $\Omega_\Lambda = 0$,
and well away from $\Omega_M = 0.3$.  
The black dashed line illustrates
how a systematic error that is proportional to $z$ diverges from
agreement with an $(0.3,0.7)$ cosmology for $z>0.5$.   \label{hubble} }
\end{figure}

The key evidence on a value for $H_0\,t_0$ comes from Figure 4.
All the supernova and SBF distances are
consistent at low redshift, and the supernova data indicate that a
model with $\Omega_\Lambda$=0 does not fit at z$\sim$ 0.5.  This is
the case for an accelerating universe, whether urged on by a
cosmological constant or by something which varies with time.  The
best fit model has $\Omega_M = 0.3$ and $\Omega_\Lambda = 0.7 $.  None
of the observers is satisfied with the current state of the
statistical errors (about twice as large for each SCP supernova as for
the High-Z data) or with current limits on possible systematic effects
that might make distant supernovae dimmer. 

\begin{figure}[htbp!]
\plotfiddle{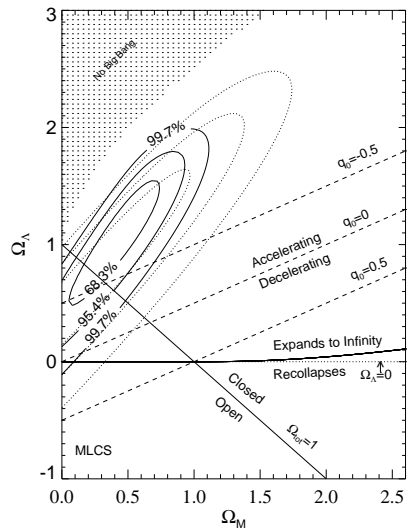}{200pt}{0}{30}{30}{-100pt}{-25pt}
\caption{ The two parameter confidence contours are
shown for the range of cosmological parameters  $\Omega_M$ and
$\Omega_\Lambda$, given the current SNIa data from the High-Z
collaboration.  The solid and dotted contours are the constraints with
and without the Fall 99 data, demonstrating the value of even a few
points at $z>0.9$.  \label{confidence} }
\end{figure}

Figure 4 shows that a
systematic effect that just grows as the redshift is not a good fit to
the data, but the most telling way to separate a systematic effect
that is proportional to redshift or time is to look at redshifts above
1.  For a $\Lambda$-dominated cosmology, there is a transition,
somewhere around $z \sim 1$, from acceleration here and now to
deceleration in the distant past when the matter density, which scales
as $(1 + z)^3$ would have been more important.  The High-Z Team is
working hard to test these ideas.  Our Fall 1999 data, which are being
slowly beaten into photometric perfection, emphasize z$\sim$1 objects
which will provide strong evidence to distinguish cosmology from
systematics.  Preliminary reductions are shown here.  Our Fall 2000
program emphasized careful UBVRI photometry in the supernova rest
frame to discern the effects of not-quite-gray dust.  The final
reduction of those data requires observations of the galaxy {\it
without} the supernova, obtained a year after the discovery, and will
be forthcoming when these templates are in hand (Jha 2001).

\section{$\Omega_M$--$\Omega_\Lambda$ constraints and the 
Distribution of $H_0\,t_0$}

We can employ the tools of likelihood analysis to construct contours
in the $\Omega_M$--$\Omega_\Lambda$ plane, as illustrated in Figure 5.
Of special note is the effectiveness of just a handful of z $>$ 0.9
supernovae in contracting the contours of this plot (as imagined by
Goobar and Perlmutter 1995).  A concise way to express the best
constraint on $\Omega_M$ and $\Omega_\Lambda$ is
$\Omega_\Lambda-1.6\,\Omega_M = 0.4\pm0.2$.  Interestingly, contours of
constant age are very nearly parallel to the long axis of the error
ellipse in Figure 5.  This means that the competition between
deceleration due to $\Omega_M$ and acceleration due to
$\Omega_\Lambda$ is well captured by the measurement of luminosity
distances, and the value of $H_0\,t_0$ is well constrained even though
the individual values of $\Omega_M$ and $\Omega_\Lambda$ are not.
 

\begin{figure}[htb]	
\plotfiddle{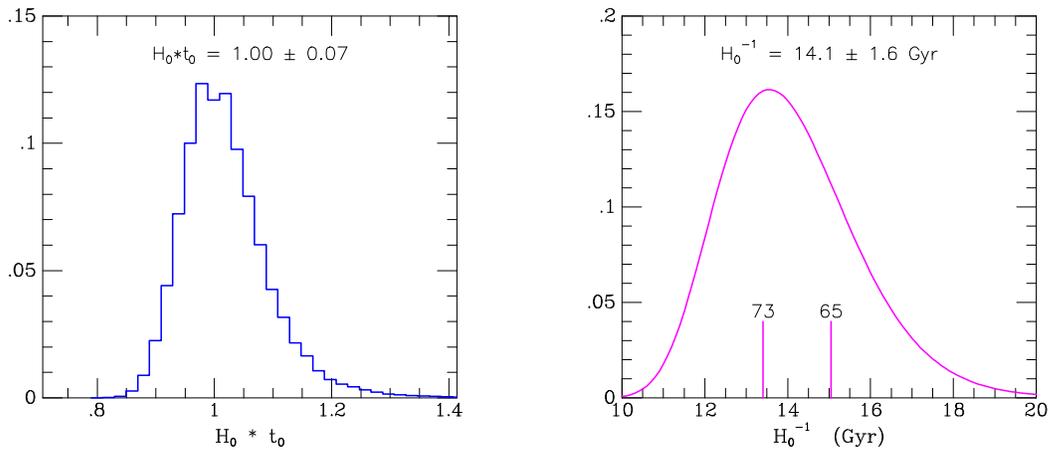}{160pt}{270}{60}{60}{-200pt}{160pt}
\caption{The left panel shows the probability density distribution for
$H_0\,t_0$ using our data.  The right panel shows the resulting
distribution for $t_0$, using the 2:1 normalization of the SNIa zero
point to $H_0 = 73$ and $65$ suggested by the HST Cepheid calibrations.
The uncertainty in  $H_0\,t_0$ ($\sim$7\%) is 
considerably smaller than the uncertainty $H_0$ ($\sim12$\%) so the
uncertainty in $t_0$ comes mainly from the $H_0$ and Cepheid zero point.
  \label{h0t0} }
\end{figure}

Marginalizing the probabilities of Figure 5 onto the $H_0\,t_0$ axis
yields a probability distribution for
$H_0\,t_0$ as shown in Figure 6.  We find $H_0\,t_0 = 1.00 \pm 0.07$. The
absolute value of the cosmic age depends on $H_0$, which, given the
excellent agreement of the SBF and SNIa distances, inherits almost
all its errors from the Cepheid zero point.
 
\section{Summary}
 
Precise distance estimators measured over the range from z$\sim$ 0 to
z$\sim$1 provide powerful constraints on the dimensionless product
$H_0\,t_0 = 1.00 \pm 0.07$.  Despite an ongoing struggle between
deceleration and acceleration, the fortuitous result is that the
formulation in elementary textbooks; $t_0 = 1/H_0$ turns out to be
accurate.  A more formal way to express the contraint from SNIa at
$z>1$ is that they imply $\Omega_\Lambda-1.6\,\Omega_M = 0.4\pm0.2$.
For our guess at the true value of $H_0$, $t_0 = 14\pm1.6$~Gyr.  A
better value for the age of the Universe hinges on nearby matters like
the distance to the LMC and metallicity dependence of Cepheids!





\acknowledgments 
Support for this work was provided by NASA through a
grant from the Space Telescope Science Institute, which is operated by
the Association of Universities for Research in Astronomy, Inc. under
NASA contract NAS5-26555.


\begin{references}
\reference Aguirre, A. 1999 \apj, 525, 583.

\reference Aguirre, A, and Haiman, Z. 2000 \apj, 532, 28.

\reference Ajhar, E.A., Tonry, J.L., Blakeslee, J.P., Riess, A.G. \& Schmidt,
        B.P. 2001, \apj, in press.

\reference Bartlett, J.G. Blanchard, A. Silk, J. and Turner, M.S. 1995 Science 267, 980.

\reference Blakeslee, J.P., Ajhar, E.A., \& Tonry, J.L. 1999, in Post-Hipparcos
        Cosmic Candles, eds.\ A. Heck \& F. Caputo (Boston: Kluwer Academic
        Publishers), 181.

\reference Coil, A. et al. 2000, \apjlett, 544, L111.

\reference Ferrarese, L.\ et al.\ 2000, \apj, 529, 745.

\reference Filippenko, A. and Riess, A.G. (1998) Phys.Rept. 307, 31-44.

\reference Freedman, W.\ L., et al. 2001, \apj, in press (astro-ph/0012376).

\reference Garnavich, P. M., et al. 1998, ApJ, 493, 53.

\reference Goobar, A. $\&$ Perlmutter, S. 1995 \apj, 450, 14.

\reference Gibson, B. K. $\&$ Stetson, P.B. 2001 \apjlett, 547, L103.

\reference Hamuy,M. et al. 1996 \aj, 112, 2398.

\reference Jha, S. et al.  1999 \apjsupp, 125, 73.

\reference Jha, S. et al. 2001 astro-ph/0101521.

\reference Leibundgut, B. et al. 1996 \apjlett, 466, 21.

\reference Norgaard-Neilsen et al. 1989 Nature 339, 523.

\reference Parodi, B.R., Saha, A., Sandage, A., \& Tammann, G.A. 2000, \apj, 540,  634.

\reference Perlmutter et al. 1995, \apjlett, 44, L41.

\reference Perlmutter et al. 1997, \apj, 483, 565.

\reference Perlmutter et al. 1998,  Nature 391, 54.

\reference Perlmutter et al. 1999, \apj, 517, 565.

\reference Phillips, M. M. 1993, \apj, 413, L105.

\reference Riess, A. G., Press, W. H., \& Kirshner, R. P. 1996, \apj, 473, 88.

\reference Riess et al. 1998 \aj, 116, 1009.

\reference Riess et al. 2000 \apj, 536, 62.

\reference Riess et al. 2001 astro-ph/0104455.

\reference Saha, A. et al. 1997, \apj, 486, 1.
 
\reference Schmidt et al. 1998 \apj, 507, 46.



\end{references}
\end{document}